\documentclass[twocolumn,prb,amsmath,amssymb,longbibliography,floatfix,superscriptaddress,preprintnumbers,nofootinbib]{revtex4-2}
\usepackage{amsmath}
\usepackage{placeins}
\usepackage{amssymb}
\usepackage{color}
\usepackage{comment}
\usepackage{physics}
\usepackage{graphicx}
\usepackage{cancel}
\usepackage{tabularx}
\usepackage{dcolumn} 
\usepackage{bm}      
\usepackage{array}   
\usepackage{amsfonts}
\usepackage{url}
\usepackage{endnotes}
\usepackage[bookmarks=false,colorlinks,citecolor=red]{hyperref}
\usepackage{ulem}
\usepackage{placeins} 

\usepackage[color=green!60]{todonotes}
\usepackage[utf8]{inputenc}
\usepackage[T1]{fontenc}

\DeclareUnicodeCharacter{2212}{-}

\begin{document}

\preprint{\large SAND2024-16729O}

\include{MyCommand}
\newcommand{\BR}[1]{{\color{red}{#1}}}
\newcommand{\AL}[1]{{\color{blue}{#1}}}
\newcommand{\JK}[1]{{\color{green}{#1}}}
\newcommand{\JA}[1]{{\color{cyan}{#1}}}
\newcommand{\CM}[1]{{\color{orange}{#1}}}

\newcommand{\BRcom}[1]{\textcolor{red}{\textbf{[BR: #1]}}}
\newcommand{\BRadd}[1]{{\color{red}#1}}

\newcommand{\ALcom}[1]{\textcolor{blue}{\textbf{[AL: #1]}}}
\newcommand{\ALadd}[1]{{\color{blue}#1}}

\newcommand{\JAcom}[1]{\textcolor{cyan}{\textbf{[JA: #1]}}}
\newcommand{\JAadd}[1]{{\color{cyan}#1}}
\newcommand{\JArem}[1]{{\color{cyan}\sout{#1}}}

\newcommand{\jkcom}[1]{\textcolor{magenta}{\textbf{[JK: #1]}}}
\newcommand{\jkadd}[1]{{\color{magenta}#1}}
\newcommand{\jkrem}[1]{{\color{magenta}\sout{#1}}}

\newcommand{\CMcom}[1]{\textcolor{orange}{\textbf{[CM: #1]}}}

\newcommand{\xref}[1]{\textbf{\color{red}[XX: REF #1]}}
\newcommand{\xtask}[1]{\textbf{\color{red}[XX: #1]}}

\title{Identifying Band Inversions in Topological Materials Using Diffusion Monte Carlo}

\author{Annette Lopez}
 \affiliation{Department of Physics, Brown University, Providence, RI 02912, United States
 }

 \author{Cody A. Melton}
 \affiliation{High Energy Density Physics Theory, Sandia National Laboratories, Albuquerque, NM 87123, United States}

 \author{Jeonghwan Ahn}
 \affiliation{Materials Science and Technology Division, Oak Ridge National Laboratory, Oak Ridge, TN 37831, United States}
 
\author{Brenda M. Rubenstein}
 \affiliation{Department of Chemistry, Brown University, Providence, RI 02912
 }
  \affiliation{Department of Physics, Brown University, Providence, RI 02912, United States
 }

  \author{Jaron T. Krogel}
  \email{Author to whom correspondence should be addressed: krogeljt@ornl.gov.}
 \affiliation{Materials Science and Technology Division, Oak Ridge National Laboratory, Oak Ridge, TN 37831, United States}

\date{\today}
\begin{abstract}
Topological insulators are characterized by insulating bulk states and robust metallic surface states. Band inversion is a hallmark of topological insulators: at time-reversal invariant points in the Brillouin zone, spin-orbit coupling (SOC) induces a swapping of orbital character at the bulk band edges. Reliably detecting band inversion in solid-state systems with many-body methods would aid in identifying possible candidates for spintronics and quantum computing applications and improve our understanding of the physics behind topologically-nontrivial systems. Density functional theory (DFT) methods are a well-established means of investigating these interesting materials due to their favorable balance of computational cost and accuracy, but often struggle to accurately model the electron-electron correlations present in the many materials containing heavier elements. In this work, we develop a novel method to detect band inversion within continuum quantum Monte Carlo (QMC) methods that can accurately treat the electron correlation and spin-orbit coupling crucial to the physics of topological insulators. Our approach applies a momentum-space-resolved atomic population analysis throughout the first Brillouin zone utilizing the Löwdin method and the one-body reduced density matrix produced with Diffusion Monte Carlo (DMC). 
We integrate this method
into QMCPACK, an open source \textit{ab initio} QMC package, so that these ground state methods can be used to complement experimental studies and validate prior DFT work on predicting the band structures of correlated topological insulators. We demonstrate this new technique on the topological insulator bismuth telluride, which displays band inversion between its Bi-$p$ and Te-$p$ states at the $\Gamma$-point. We show an increase in charge on the bismuth $p$ orbital and a decrease in charge on the tellurium $p$ orbital when comparing band structures with and without SOC. Additionally, we use our method to compare the degree of band inversion present in monolayer Bi$_2$Te$_3$, which has no interlayer van der Waals interactions, to that seen in the bulk. The method presented here will enable future, many-body studies of band inversion that can shed light on the delicate interplay between correlation and topology in correlated topological materials. 
\end{abstract}

\keywords{band inversion, Diffusion Monte Carlo, topology, electron correlation, bismuth telluride, L{\"o}wdin population analysis}

\maketitle
\maketitle

\footnote{This manuscript has been authored by UT-Battelle, LLC under Contract No. 
DE-AC05-00OR22725 with the U.S. Department of Energy. The United States 
Government retains and the publisher, by accepting the article for publication, 
acknowledges that the United States Government retains a non-exclusive, 
paid-up, irrevocable, worldwide license to publish or reproduce the published 
form of this manuscript, or allow others to do so, for United States Government 
purposes. The Department of Energy will provide public access to these results 
of federally sponsored research in accordance with the DOE Public Access Plan 
(http://energy.gov/downloads/doe-public-access-plan).}


\section{Introduction}

Topological insulators (TIs) are a fascinating - and rapidly growing - class of quantum materials somewhat paradoxically characterized by insulating bulk, yet metallic surface states.  Unlike those in conventional, non-topological insulators, the surface states of TIs are symmetry-protected and thus unusually robust to perturbations or defects. This gives rise to an array of intriguing quantum phenomena including the Quantum Spin Hall (QSH) \cite{kane_2005, bernevig_2006} and Quantum Anomalous Hall (QAH) effects \cite{haldane_1983, onoda_2003, chang_2013}. 

The unusual surface states in TIs stem from their twisted band structures, which distinguish them from conventional insulators and enable a hallmark of TIs: band inversion. At time-reversal-invariant-momentum (TRIM) points in TIs' Brillouin zones, spin-orbit coupling (SOC) induces a swapping of orbital character at the bulk band edges. As a result, these materials' valence bands inherit the orbital character of their conduction bands and vice-versa \cite{anirban_2023, kane_2011, narang_2021, bansil_2016}. One well-studied class of structurally-similar 3D topological insulators that exhibit band inversion consists of Bi$_2$Se$_3$, Bi$_2$Te$_3$, and Sb$_2$Te$_3$ \cite{zhang_2009, zhao_2015}. All of these materials display band inversion between the $p$-state conduction band minimum and valence band maximum at the $\Gamma$ point \cite{zhang_2009}. 

Because of the primacy of these materials' band structures for describing their properties, they are often modeled using Density Functional Theory (DFT) \cite{zhang_2013, hao_2019, vidal_2011}, which can readily compute band structures with reasonable accuracy for weakly-correlated materials. While a DFT description suffices for many topological insulators, there are an increasing number of TIs that exhibit stronger electron correlation and hence are not as accurately described by a mean-field DFT picture. One such example is SmB$_6$, a three-dimensional strongly-correlated topological Kondo insulator that uniquely displays band inversion between the Sm-5$d$ and Sm-4$f$ orbitals. \cite{li_2020, sakhya_2020, neupane_2013}. While one would often turn to many-body methods that can naturally account for correlation to model such correlated TIs, many such methods, including quantum Monte Carlo (QMC) methods \cite{metropolis_1949, kalos_1962, mcmillan_1965, anderson_1975, anderson_1976, ceperley_1980, }, cannot readily provide band structure information. For this class of high-interest materials, it would thus be beneficial to have techniques that not only characterize aspects of their topology, but also capture their stronger electron-electron interactions with sufficient accuracy.

In this manuscript, we attempt to bridge this gap by developing a new means of identifying topological insulators based upon their band inversion within Diffusion Monte Carlo (DMC), a continuum flavor of QMC. The crux of our idea is that, even though DMC cannot provide full band structures, it can characterize the bands directly surrounding the Fermi level. From this information, one can identify a characteristic of topological insulators within DMC by determining if the orbital character of the bands has been inverted. Given the growing number of potentially strongly correlated topological insulators discovered in recent years \cite{watanabe_2019, yujun_2020, deng_2021, liu_2023}, a many-body approach which can capture the effect of correlation on topological properties will be increasingly valuable. More broadly, the formalism we present here can also be employed by a wide range of other wave function theories, facilitating the application of many-body methods to topological materials.

Using Bi$_2$Te$_3$, a paradigmatic topological insulator as an example, we show how the orbital character of bands generated in DMC can be resolved using a L\"{o}wdin transformation of the occupied bands into their constituent atomic orbitals. Then, employing recently developed techniques for incorporating spin-orbit coupling into DMC calculations  \cite{melton_2016, melton_2016_2, melton_2017}, we demonstrate that DMC can detect the band inversion expected in Bi$_2$Te$_3$ when spin-orbit coupling is present, but not in its absence, which serves as a strong check on our method. Lastly, as a practical application of our approach in a novel setting, we use our method to compare the strengths of band inversion in monolayer and bulk Bi$_2$Te$_3$, finding that the monolayer does not display a band inversion signal at the $\Gamma$-point, in direct contrast with the bulk. The techniques set forth here provide a path toward studying topology in correlated topological materials using Diffusion Monte Carlo and other many-body methods.

Below, we first detail the approach we have developed to resolve the atomic orbital character of the occupied bands in Section \ref{sec:methods}. Then, we describe how we perform Diffusion Monte Carlo calculations with spin-orbit coupling in Section \ref{sec:comp_details}. This new capability is demonstrated by analyzing band inversion in 3D and monolayer Bi$_2$Te$_3$ in Section \ref{sec:results}. We conclude with a discussion of the potential applications and extensions of our method in Section \ref{conclusions}. 

\section{Method: Obtaining Band Inversions from Atomically-Projected One-Body Reduced Density Matrices}
\label{sec:methods}

In order to track potential band inversions involving bands with different orbital characters, we first re-express our 1-body reduced density matrices (1RDMs), which are derived from the many-body wavefunction and initially written in terms of momentum-dependent Bloch states, into atomic orbitals. To do so, we perform an orthogonal transformation of the DMC 1RDM.

In a crystalline system, the 1RDM can be
expressed in an arbitrary single-particle basis of Bloch spinors, ${\phi_{ki}}$, that can be explicitly momentum-dependent such that
\begin{equation}
\label{eq:main_eqn}
\hat{n}_{1} = \sum_k \sum_{ij}|\phi_{ki}\rangle n_{kij}\langle\phi_{kj}|,
\end{equation}
where $k$ is the crystal momentum ($k$-point) of each Bloch spinor and the indices $i$ and $j$ range over all spinors at each respective $k$-point. The 1RDM is essentially collection of number operators $n_{kij}$, which are partitioned by momenta, $k$, located within the first Brillouin Zone. For the purpose of identifying band inversion at specific high-symmetry points in the BZ, we choose not to sum over the BZ in Equation \ref{eq:main_eqn} to obtain the k-dependent density matrix

\begin{equation}
    \hat{n}_k = \sum_{ij} |\phi_{ki}\rangle n_{kij}\langle\phi_{kj}|. 
\label{eq:k_main_eqn}
\end{equation}

For an insulating system (including TIs), the trace of each $k$-resolved density matrix is $\mathrm{Tr}(\hat{n}_k)=N_{\mathrm{prim}}$, the number of electrons per primitive cell.  With a further partitioning into atomic contributions, we can resolve the twisting effect due to SOC on the occupied bands. Thus, with $\hat{n}_k$ in hand, we 
perform a transformation to an atomic orbital (AO) basis to extract specific occupation values at $k$-points in the Brillouin zone at which band inversion is suspected to occur. 

Although the atomic partitioning is not unique, and in some sense is a free choice, we only require a suitable partitioning capable of providing a reliable indicator of the inter-atomic occupancy exchange that defines band inversions. Taking inspiration from the L\"owdin population analysis \cite{lowdin_1950, lowdin_1970}, we switch to a more local description by projecting the Bloch states onto an orthogonal atomic orbital basis, $\{\psi_l^a\}$, through the use of an overlap matrix
\begin{equation}
S_{ki}^\ell= \langle \psi_{\ell}^a|\phi_{ki}\rangle.
\label{eq:overlap}
\end{equation}
Here, $a$ labels an atom, while $\ell$ denotes the angular state, i.e., $\ell\in \{s,p,d,f\}$.  Orthogonal AOs, can be constructed via the L\"owdin symmetric orthogonalization procedure \cite{lowdin_1950, lowdin_1970}. In this work, we utilize a feature which projects wavefunctions onto orthogonalized atomic wavefunctions in Quantum Espresso to facilitate our method. During this process, we anticipate a small partial loss of charge due to spillage, the error incurred from an incomplete basis set while projecting onto an atomic basis \cite{shanchez_portal_1996}. With these projections, we can transform the 1RDM expressed in terms of Bloch states into a 1RDM expressed in terms of orthogonal atomic states. To do so, we insert the identity operator resolved in the same atomic basis, $1 = \sum_{\ell} |\psi_{\ell}^a\rangle \langle \psi_{\ell}^a|$, on either side of $n_{kij}$ in Equation \ref{eq:k_main_eqn} to obtain
\begin{equation}
\hat{n}_k =\sum_{ll'}|\psi_l^a\rangle\sum_{ij}S^\ell_{ki}n_{kij}S^{\ell'*}_{kj}\langle\psi_{\ell'}^a|.
\label{transformed}
\end{equation}
For each atom's set of orbitals, we then obtain a per-$k$-point occupation that quantifies the fractional occupation of the $\ell$th orbital of atom $a$ 
\begin{equation}
    \label{eq:k_atomic_occ}
    N_{a\ell}(k) = \sum_{o\in \ell}\langle \psi_o^a|\hat{n}_k|\psi_o^a\rangle = \sum_{o\in \ell}\sum_{ij}S^o_{ki}n_{kij}S^{o*}_{kj}, 
\end{equation}
where $o \in \{\{s\}, \{p_x,p_y,p_z\}, \{d_{xy}, d_{yz}, d_{xz}, d_{x^2-y^2}, d_{z^2}\},\ldots\}$. The Löwdin charges are projected onto the $l$ orbitals in both the non-SOC and SOC cases.

Band inversion can then be detected using $N_{a\ell}(k)$ by observing a loss of weight/occupancy from one site-resolved atomic orbital and a corresponding gain in another, \textit{e.g.}, in the neighborhood of TRIM points in the first Brillouin zone.  In particular, to resolve SOC-driven band inversions most clearly, we use the non-SOC case as a reference to isolate SOC-induced changes in the occupations.  We thus use 
\begin{equation}
\label{eq:k_atomic_occ_diff}
\delta N_{a\ell}(k)\equiv N_{a\ell}^{\mathrm{SOC}}(k)-N_{a\ell}^{\mathrm{no-SOC}}(k)
\end{equation}
as the primary quantity to assess the presence or absence of band inversions in the target material.

\section{Computational Details}
\label{sec:comp_details}
\subsection{Bismuth Telluride (Bi$_2$Te$_3$)}
\label{sec:materials}
\begin{figure}[hbt!]
\includegraphics[width=\columnwidth]{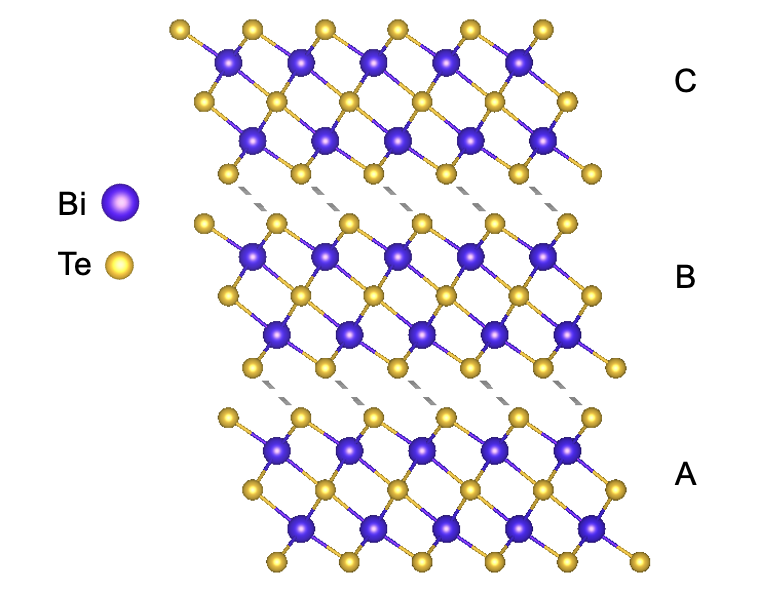}
\caption{The crystal structure of bulk Bi$_2$Te$_3$. The letters A, B, and C denote the layers in an ABC-stacking configuration.}
\label{ex}
\end{figure}
For illustrative purposes, we choose a well-characterized topological material, bismuth telluride (Bi$_2$Te$_3$) \cite{witting_2019}. Bulk Bi$_2$Te$_3$ has a rhombohedral crystal structure and is composed of sets of hexagonal close-packed quintuple layers. These sets of five layers follow the sequence Te(top)-Bi-Te(middle)-Bi-Te(bottom). The material displays ABC stacking with consecutive Te layers weakly bound by van der Waals interactions. Bismuth telluride is subject to strong spin-orbit interactions which lead to band inversion between the Bi-$p$ and Te-$p$ states at the $\Gamma$-point; SOC effects make the energies of the Bi-$p$ states lower than those of the Te-$p$ states \cite{zhang_2009}. The SOC-enabled orbital hybridization through the inversion opens up a gap on the order of approximately $0.1$ eV \cite{chen_2009, michiardi_2014, nechaev_2013, noh_2008, kioupakis_2010}. 

The crystal structure for bulk Bi$_2$Te$_3$ used in this work was taken from the Inorganic Crystal Structure Database (ICSD, accessed 2023) \cite{bergerhoff_1983}. The monolayer structure was extracted from the bulk structure and then modeled using a $23$ \r{A} vacuum (see Supporting Information).

\subsection{Density Functional Theory Calculations}
\label{sec:DFT_calcs}

We first utilized Kohn-Sham DFT \cite{kohn_1965} to obtain single-particle Bloch spinors for subsequent DMC calculations. For this purpose, we used the Perdew-Burke-Ernzerhof (PBE) \cite{perdew_1996} exchange-correlation functional, as implemented in the Quantum Espresso code (v.7.0) \cite{giannozzi_2009}. For bulk Bi$_2$Te$_3$, a $16 \cross 16 \cross 4$ Monkhorst-Pack \cite{monkhorst_1976} k-point grid was used to sample the Brillouin zone of the non-SOC-inclusive five-atom primitive cell, while a smaller, but sufficiently converged, $14 \cross 14 \cross 4$ grid was used in the SOC case. For monolayer Bi$_2$Te$_3$, we employed a $16 \cross 16 \cross 4$ grid for both the SOC and non-SOC calculations. The DFT calculations were also used to obtain the orthogonalized AOs and related Bloch state-AO overlap matrix given by Equation \ref{eq:overlap} to support the atomic projection of the DMC 1RDM as shown in Eq. \ref{eq:k_atomic_occ}.

Our electronic structure calculations were performed with correlation consistent effective core potentials (ccECPs) \cite{bennett_2017}. By construction, the effective potentials efficiently simulate the interactions between the atomic core and valence states and offer accurate spin-orbit relativistic interactions for many-body calculations. ECPs were utilized for the Bi and Te atoms with an Xe-core and a Kr-core, respectively, and a plane-wave kinetic energy cutoff of 500 Ry \cite{wang_2022}.

\subsection{Spin-Orbit Diffusion Monte Carlo Calculations}
\label{sec:dmc_calcs}
To more thoroughly account for many-body effects in our modeling, we further simulated Bi$_2$Te$_3$ using DMC, a real-space, many-body electronic structure method. DMC models the ground state of a system by solving the imaginary-time many-body Schrodinger equation 
\begin{equation}
    -\partial_{t} \Psi({\bf{R}}, t) = (\hat{H}- E_T)\Psi({\bf{R}}, t).
    \label{schrodinger}
\end{equation}
Here, $\Psi({\bf{R}}, t)$ represents the wavefunction of the system as it evolves with imaginary time, $t$; ${\bf{R}}$ represents the vector of electron coordinates; and $E_T$ represents the trial energy, the expectation value of the Hamiltonian, $\hat{H}$, with the trial wavefunction ($\Psi_T$) that is an approximation to the ground state energy and serves as an energy offset in the equation above. 

We can recast the imaginary time Schr\"{o}dinger equation into integral form by incorporating the use of a Green’s function to show that, by adjusting the trial energy, we can project out the ground state via dampening of the higher-energy states with increasing imaginary time \cite{foulkes_2001}.  For simplicity, we show the imaginary time evolution without the similarity transformation used for importance sampling:
\begin{equation}
    \Psi({\bf{R}}, t+\tau) = \int d{\bf{R}'} G({\bf{R}} , {\bf{R'}}, \tau)\Psi({\bf{R'}}, t)
\end{equation}
\begin{equation}
    G({\bf{R}} , {\bf{R}'}, \tau) = \sum_i \Psi_i({\bf{R}})e^{-\tau(E_i-E_T)}\Psi_i^*({\bf{R'}})
    \label{green}
\end{equation}
The Green's function in Equation \ref{green} also follows Equation \ref{schrodinger}, where $\{\Psi_i\}$ and $\{E_i\}$ are the complete set of eigenstates and eigenvalues of $\hat{H}$ and $\tau$ denotes the imaginary time slice \cite{foulkes_2001}. The diffusion of the trial wavefunction in imaginary time converges to the true ground state, $\Psi_0$, with ground state energy $E_0$ in the long time limit of the evolution from the initial state (\textit{i.e.}, from the trial wavefunction $\Psi_T(R)=\Psi(R,t=0)$):
\begin{equation}
    \ket{\Psi_0} = \lim_{t\rightarrow\infty}\ket{\Psi_t} = \lim_{N\rightarrow\infty}\hat{G}_\tau^N \ket{\Psi_T},
\end{equation}
where $G({\bf{R}} , {\bf{R}'}, \tau) = \langle {\bf{R}} | \hat{G}_\tau |{\bf{R}'}\rangle$. The fixed-node (FN) approximation is applied to handle fermion antisymmetry \cite{anderson_1976}, in which case the energy of the equilibrium state remains strictly an upper bound to the true ground state energy. 

DMC simulations often use a compact representation of the trial wavefunction of the Slater-Jastrow type, which multiplies spin-resolved Slater determinants by a symmetric, non-negative Jastrow factor \cite{foulkes_2001}:
\begin{equation}
    \Psi_T({\bf{R}}) = e^{J({\bf{R}})} \mathrm{det}[\{\phi_i^\uparrow({\bf{r_j^\uparrow}})\}] \mathrm{det}[\{\phi_i^\downarrow({\bf{r_j^\downarrow}})\}].
\end{equation}
In the presence of SOC, the spin no longer commutes with the Hamiltonian. That is, particle spins now enter as dynamical variables, in addition to the spatial coordinates.
To model these effects in DMC, a few important modifications to the basic DMC algorithm are made 
\cite{melton_2016, melton_2016_2, melton_2017}.
First, a continuous representation of the spin states is introduced in order to enable Monte Carlo sampling,
${\bf{R}}\in \mathbb{R}^{3N} \Rightarrow {\bf{X}} \in \mathbb{R}^{4N}$. Therefore, we move from single-particle orbitals $(\phi_i)$ to single-particle spinors $(\psi_i)$ in the Slater-determinant:
\begin{equation}
    \Psi_T({\bf{X}}) = e^{J({\bf{R}})} \text{det}[\{\psi_i({\bf{r}}_j, s_j)\}]
\end{equation}
\begin{equation}
    \psi_i({\bf{r}}_j, s_j) = \phi_i^\uparrow({\bf{r}}_j)e^{is_j} + \phi_i^\downarrow({\bf{r}}_j)e^{-is_j}
\end{equation}
With this transition, the spinors are inherently complex, 
requiring the use of
the fixed-phase approximation \cite{ortiz_1993}, even for wavefunctions with full periodic spatial symmetry: $\Psi_T({\bf{X}}) = \rho({\bf{X}})\text{exp}[i\Phi({\bf{X}})]$, $\Phi({\bf{X}}, \tau) = \Phi({\bf{X}})$.

\begin{figure*}[hbt!]
\includegraphics[width=\linewidth]{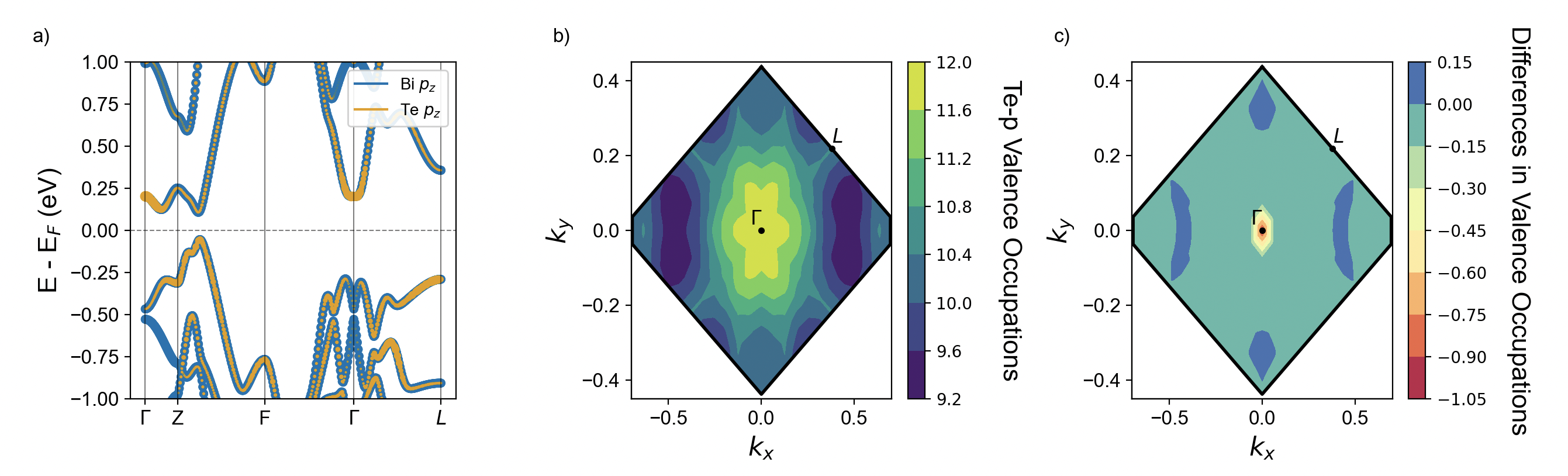}
\caption{(a) The Bi$_2$Te$_3$ PBE band structure with bands with Bi and Te $p_z$ character appearing in blue and gold, respectively. (b) As given by Equation \ref{eq:k_atomic_occ}, the total valence band PBE Te-p occupancies around the $\Gamma$-point in the $k_z=0$ plane, in the presence of SOC. (c) Difference in total valence band PBE Te-p occupancy (Equation \ref{eq:k_atomic_occ_diff}) between calculations with and without SOC.  A large change of about -0.9$e$ in the valence occupation reveals the inversion of the Bi-p and Te-p bands near the $\Gamma$-point.}
\label{dft}
\end{figure*}

Both the Variational Monte Carlo (VMC) and DMC algorithms are modified to incorporate the sampling of the spin degree of freedom, in a similar way as the spatial variables. Practically, the spin-orbit contribution to the Hamiltonian is introduced through the relativistic effective core potentials and the introduction of a fictitious spin mass sets the rate of sampling of the spins relative to the spatial diffusion timestep \cite{melton_2016, melton_2016_2, melton_2017}.

\begin{equation}
    G({\bf{X}}, {\bf{X'}}, \tau) \propto G({\bf{R}}, {\bf{R'}}, \tau)\text{exp}(-|{\bf{S}} - {\bf{S'}} - \tau_s{\bf{v}}_s({\bf{S'}})|^2/2\tau_s).
\end{equation}

Utilizing the spin-orbit-enabled algorithm described above, we collected results for several flavors of QMC, performed at $40$ k-points along the $F-\Gamma-L$ path with the QMCPACK \cite{kim_2018, kent_2020} simulation code (v3.17.9), supported by the Nexus \cite{krogel_2016} workflow automation system. We modified QMCPACK to calculate the 1RDM in the spinor basis (Equations \ref{eq:main_eqn}-\ref{eq:k_main_eqn}) and made the changes available in the public branch of the code on GitHub. To check our methodology, we additionally performed VMC calculations without Jastrow factors using  DFT-generated orbitals to compare against the DFT results, since the band inversion should be identical with a fixed Slater-determinant wavefunction, with or without stochasticity in VMC and DFT, respectively.

Following this validation step, we then proceeded with a traditional QMC construction of the trial wavefunction using a Slater determinant that incorporates the electronic correlation for the DMC calculations via one- and two-body Jastrow factors. The one-body term models the electron-nuclear interactions and the two-body term models the electron-electron correlations. VMC optimization of the Jastrow factors was performed at the $\Gamma$-point via the linear optimization method \cite{toulouse_2007}. The optimization was judged to have acceptable accuracy, achieving a variance-to-energy ratio of 0.02 Ha. In contrast with similar solid-state QMC calculations, we do not converge our results with respect to one- and two-body finite-size effects. We instead focus on demonstrating our methodology, which can be extended to handle finite-size effects in future works. DMC calculations in twisted boundary conditions \cite{lin_2001} were then performed at each $k$-point along the $F-\Gamma-L$ path.  A direct output of the QMC calculations is the $n_{kij}$ matrix (Equation \ref{eq:k_main_eqn}) that was subsequently used to obtain $N_{a \ell}(k)$ (Equation \ref{eq:k_atomic_occ}) to characterize the band inversion.

\section{Results}
\label{sec:results}

\subsection{Characteristics of Bulk Bi$_2$Te$_3$ Band Inversion in DFT}

Before using our new DMC approach to study band inversion, we begin by studying the characteristics of the inversion using DFT. We do so to both orient the reader regarding key features of the band inversion and to establish a set of DFT reference predictions for this weakly correlated material.  In Figure \ref{dft}(a), we plot the PBE band structure of bulk Bi$_2$Te$_3$ with SOC across a high-symmetry k-path. Bi-$p_z$ orbital character is indicated in blue, while Te-$p_z$ character is indicated in gold. As can be seen in the figure, the valence bands below the Fermi energy generally have Te-$p_z$ character across the $k$-path while the conduction bands above the Fermi energy have Bi-$p_z$ character. As expected based on previous literature \cite{zhang_2009},
one can see an inversion of these colors near the $\Gamma$-point, which is indicative of a band inversion between the Te-$p$ and Bi-$p$ orbitals in this region of $k$-space.

To provide more perspective, we also plot a 2D heat map of the occupancy of the valence band across the Brillouin zone in the presence of SOC in Figure \ref{dft}(b). Such heat maps will prove particularly useful in our subsequent discussion because, while DFT can generate full band structures as in Figure \ref{dft}(a), current DMC methods can only analyze the energies and occupancies of the valence band, as this portion of the band structure corresponds to the ground state wavefunction. We see from this heat map that the largest Te-$p$ occupancies (Equation \ref{eq:k_atomic_occ}) occur around the $\Gamma$-point, consistent with the valence band structure in Figure \ref{dft}(a). We present an additional DFT analysis of the L\"owdin charges for the individual Bi and Te atoms in Supplemental Information Tables S3 and S4. The Te-$p$ and Bi-$p$ occupation values presented in this work are a sum of those from the three Te atoms and the two Bi atoms found in the unit cell, respectively. A breakdown of the individual atomic contributions in the unit cell reveals that the middle Te layer contributes roughly 3.8 electrons to the Te-$p$ orbital, while the outer two tellurium layers contribute about 3.7 (3.6) electrons without (with) SOC. We observe that charge flows from the $p$-orbitals of the outer tellurium layers to the $p$-orbitals of the bismuth atoms when SOC is turned on (compare Tables S3 and S4). 

To see the SOC effects on the Te-p occupancies, we can take the difference between 2D slices of the Brillouin zone, as in Figure \ref{dft}(b), for the system with and without SOC (Equation \ref{eq:k_atomic_occ_diff}). In Figure \ref{dft}(c) at the $\Gamma$-point, we see a clear difference in the Te-$p$ occupancies on the order of -1 (nearly complete deoccupation of a single state), marking a significant decrease in the occupancy in the presence of SOC. Thus, even though the Te-$p$ occupation appears large in Figure \ref{dft}(b) when SOC is present, we see that incorporating SOC results in a net loss of nearly unit Te-$p$ occupation at the $\Gamma$-point. From this figure, we can also observe that the Te-$p$ orbitals gain occupation at other regions of the Brillouin zone. This is consistent with the band structure depicted in Figure \ref{dft}(a). Altogether, these plots demonstrate that band inversions can be identified based on ground state information alone, namely, valence band occupations. This is a key point that makes DMC studies of band inversion possible in the first place.

\subsection{Method Validation: Consistency Between DFT and Slater-Only VMC}

Now that we have demonstrated the utility of using valence band occupations to identify band inversions, we now turn to looking for these same signatures using valence band occupations produced by QMC. In contrast with the DFT occupations, the QMC occupations will reflect many-body effects.

\begin{figure}[hbt!]
\includegraphics[width=\columnwidth]{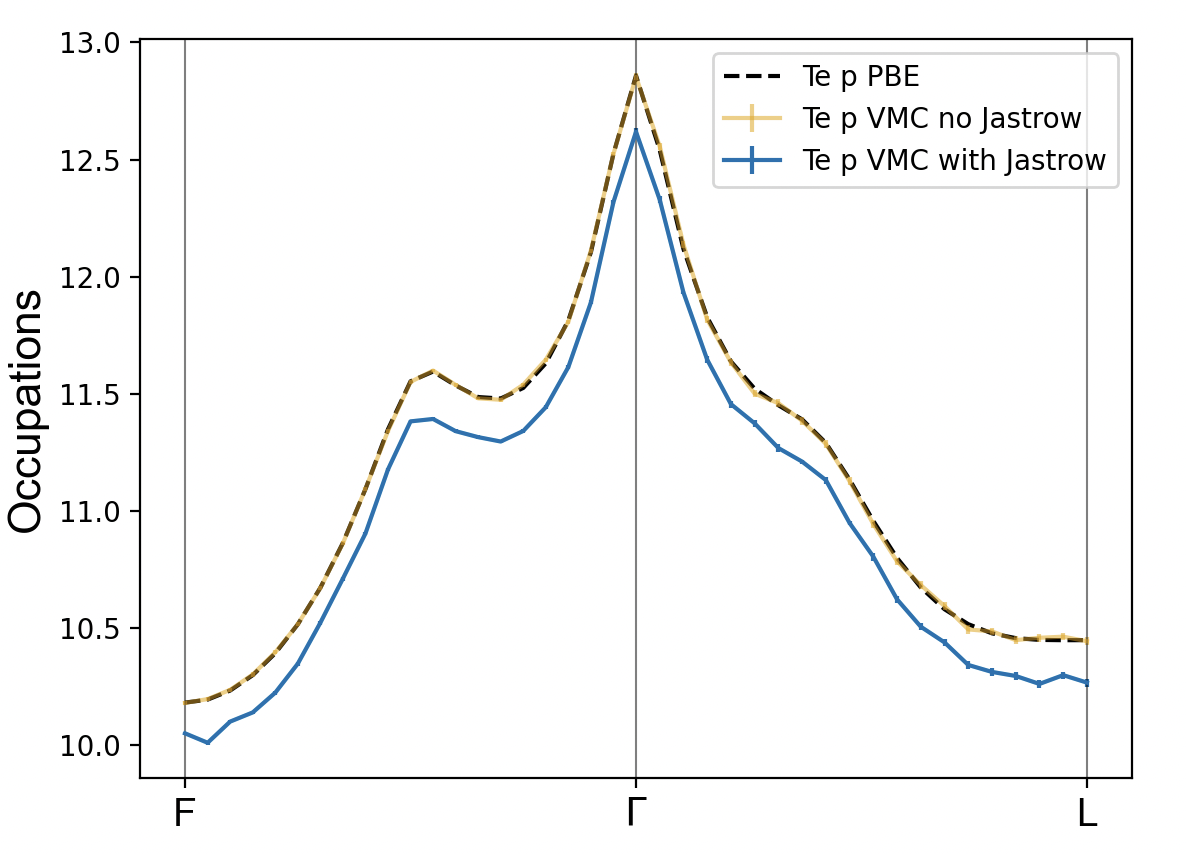}
\caption{Comparison of total Te$-p$ occupancies along the $F-\Gamma-L$ path without SOC for different levels of theory (DFT and VMC).  As required, PBE and Jastrow-free VMC occupations exactly match. The introduction of explicit dynamic correlation in VMC shifts the occupancies relative to the PBE initial state. 
}
\label{te}
\end{figure}

We first present the Te-$p$ occupancies along the $F-\Gamma-L$ path without SOC effects obtained using PBE, VMC without a Jastrow, and VMC with a Jastrow, as seen in Figure \ref{te}. We compare the curves obtained using these different methods to assess how the incorporation of different degrees of electron correlation will impact the occupations. Note that the $k$-path we illustrate here has been shortened relative to the DFT band structure in Figure \ref{dft}(a) due to the increased expense of QMC calculations.  We construct our high-symmetry path from a total of 40 
$k$-points, with 20 points along each of the $F-\Gamma$ and $\Gamma-L$ segments.

In Figure \ref{te}, we first compare the DFT occupancies to those obtained using VMC without a Jastrow. Since a Jastrow was not employed in the VMC, the orbitals from the prior DFT calculation remain unchanged. We find the DFT and VMC no-Jastrow results match perfectly within the available statistical resolution. This verifies that the projection onto atomic orbitals has been correctly applied to the QMC data. In agreement with Figure \ref{dft}(b), the mean of the Te-$p$ valence occupations is about 11 electrons across the Brillouin zone (see Supplemental Information Table S5). We next introduce an energy-optimized Jastrow correlation factor (limited to one- and two-body correlations) into the VMC trial wavefunction. Notably, the correlations introduced by the VMC wavefunction with Jastrow factors significantly modify the occupations in a relatively uniform fashion across the Brillouin zone. Indeed, we observe a consistent reduction in the Te-$p$ occupations with the incorporation of the Jastrow factors, resulting in roughly $0.15-0.20$ electrons being removed from the Te-$p$ states on average. 

\subsection{Signatures of Bulk Bi$_2$Te$_3$ Band Inversion in DMC}
After performing this consistency check without accounting for spin-orbit coupling, we employ fixed-node and fixed phase spin-orbit DMC 
 (Section \ref{sec:dmc_calcs}) and purposefully focus on the DFT and DMC occupations for concision for the remainder of the discussion. We note that the $S^l_{ki}$ matrices (Equation \ref{eq:overlap}) introduce a systematic error with respect to $k$. To reduce this sensitivity, we chose to limit our analysis to the diagonal elements of the 1RDM, effectively making the approximation that the natural orbitals do not change to first-order in correlation.
 
\begin{figure*}[hbt!]
\includegraphics[width=.9\linewidth]{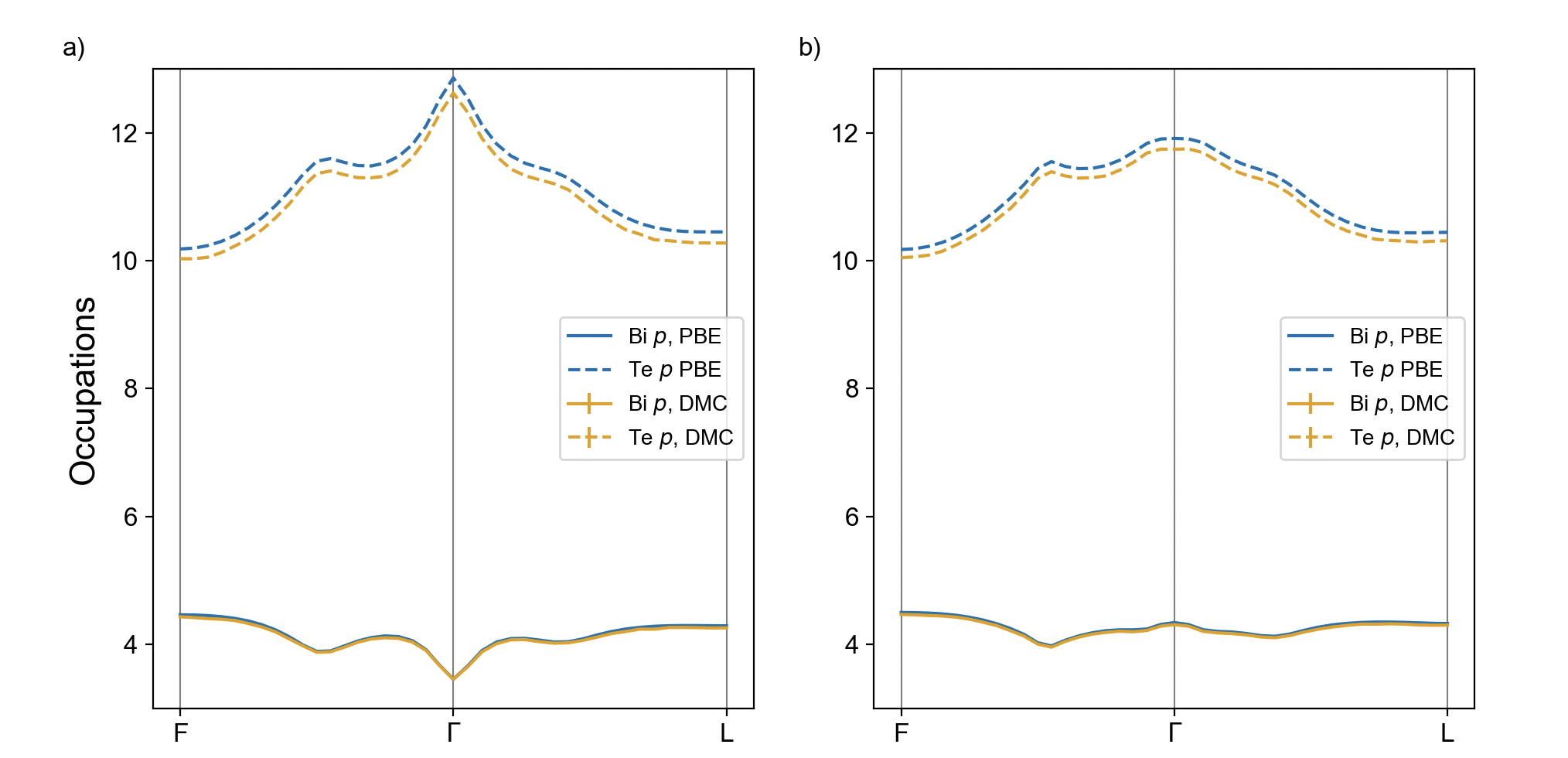}
\caption{PBE and DMC Bi and Te $p$ occupancies (Equation \ref{eq:k_atomic_occ}) along the $F-\Gamma-L$ $k$-path (a) without and (b) with SOC. }
\label{soc}
\end{figure*}
In Figure \ref{soc}, we consider how the occupations change in DFT (PBE) and DMC with and without SOC. There is relatively good agreement between the PBE and DMC valence occupations (Equation \ref{eq:k_atomic_occ}), both with and without SOC, possibly because of the relatively weak electron correlation in Bi$_2$Te$_3$. The PBE and DMC predictions of the Bi-$p$ occupations almost perfectly coincide across our $k$-path. The Te-$p$ occupation curves are also in agreement, but run parallel to one another with PBE predicting occupancies roughly 0.1 electrons greater than the DMC occupancies across the path. This is likely due to charge build up in the interstitial regions between consecutive tellurium layers where van der Waals interactions are strong. That is, PBE confines the charge to the quintuple layers, while DMC redistributes about 0.3$e$ in a nonlocal fashion (see Supplemental Information Tables S3, S4, S6, and S7), independent of $k$.

We now compare the results from the SOC and non-SOC calculations. The SOC occupations (Figure \ref{soc}(b)) exhibit pronounced deviations relative to the non-SOC occupations (Figure \ref{soc}(a)) near the $\Gamma$-point: the Bi-$p$ values significantly dip while the Te-$p$ values significantly rise. This is indicative of the band inversion; the occupation of the Bi-$p$ orbital is increased around the $\Gamma$-point as a result of charge being transferred into the orbital, while the occupation of the Te-$p$ orbital slightly decreases at the $\Gamma$-point and flattens at adjacent k-points. With SOC, some charge is thus transferred from the Te-$p$ orbital to the Bi-$p$ orbital. We have demonstrated that our technique effectively detects a transition in the occupations when SOC is turned on, consistent with the emergence of a topologically non-trivial state. 

In Figure \ref{ex}, we further examine the change in the occupations in the presence of SOC by taking the difference between the occupations with and without SOC (Equation \ref{eq:k_atomic_occ_diff}). The Te-$p$ and Bi-$p$ occupations exhibit a sharp decrease and increase at the $\Gamma$-point, respectively. These reciprocal changes in occupation by nearly a full electron are a strong indication of band inversion at the $\Gamma$-point.
This is in agreement with Figure \ref{dft}(a): upon band inversion, we expect the valence bands to inherit an increased Bi-$p$ character and the conduction bands to inherit an increased Te-$p$ character. Less significant fluctuations and changes in the occupations can be observed at other points along the $k$-path. This further corroborates that the changes in the occupations we see at the $\Gamma$-point stem from band inversion because band inversion is expected to be most significant at the $\Gamma$-point.

\begin{figure*}[hbt!]
\includegraphics[width=.9\linewidth]{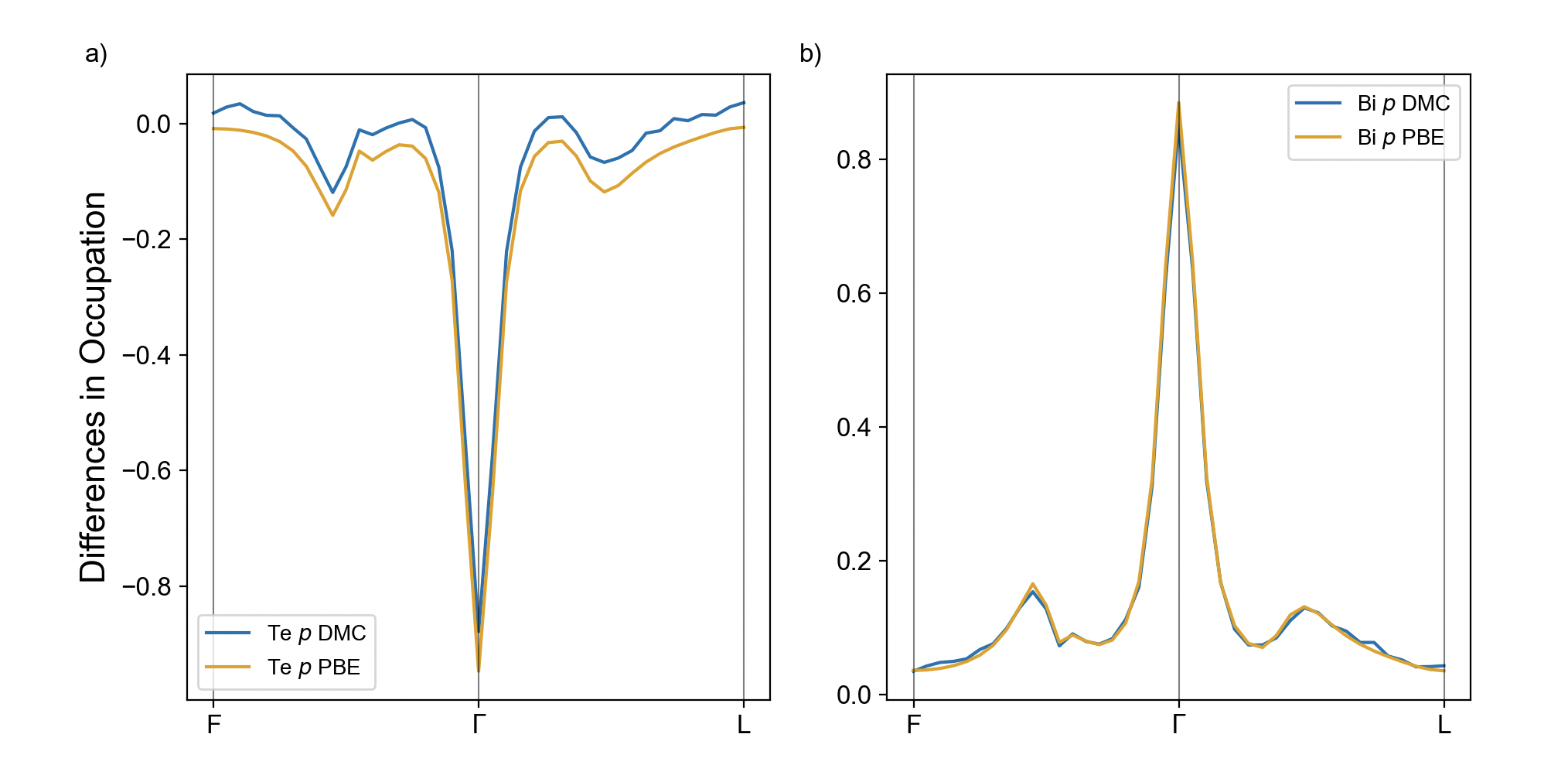}
\caption{Differences in occupations between SOC and non-SOC calculations (Equation \ref{eq:k_atomic_occ_diff}) along the $F-\Gamma-L$ path, clearly showing the band inversion between the Te-$p$ and Bi-$p$ orbitals near the $\Gamma$-point. (a) Differences between the Te-$p$ DMC and PBE occupancies. (b) Differences between the Bi-$p$ DMC and PBE occupancies.}
\label{ex}
\end{figure*}

Interestingly, the magnitudes of the Te-$p$ and Bi-$p$ signals differ. Whereas the PBE Te-$p$ occupation decreases by nearly a whole electron at the $\Gamma$-point, the PBE Bi-$p$ occupation increases by roughly 0.85 electrons, suggesting that charge is being transferred to the Te $p$ orbitals from other sources as well. Looking at the corresponding DMC Te-$p$ difference, we see the magnitude of the signal is slightly reduced, while the DMC and DFT Bi-$p$ signals remain in agreement, within the statistical uncertainty. This slight reduction in the Te-$p$ difference reflects a balancing of the occupation of the Bi-$p$ orbital and the de-occupation of the Te-$p$ orbital.

 In order to illustrate the utility of our approach, we next apply it to a system that is expected to be topologically-trivial: monolayer Bi$_2$Te$_3$. We find that the monolayer configuration, absent van der Waals effects between consecutive tellurium layers, lacks a band inversion signal at the $\Gamma$-point.
Across the path, the average change in Bi-$p$ and Te-$p$ occupations when SOC is introduced is approximately 0.1$e$ (see Supplemental Information Tables S17 and S18). At the $\Gamma$-point, there are no significant, coinciding decreases or increases in Te-$p$ and Bi-$p$ occupations, as SOC is turned on (see Supplemental Information Figure S1). The disappearance of the band inversion signal at the $\Gamma$-point in the monolayer emphasizes the role of van der Waals forces in these systems. In comparison to the bulk, we can see that these weak inter-layer interactions facilitate the coupling of electronic states and band inversion in topological materials. Our method demonstrates a loss of topological character in monolayer Bi$_2$Te$_3$ with no band inversion signal present at the $\Gamma$-point. Thus, our DMC approach concurs with the topological triviality expected in the monolayer setting due to the loss of interlayer interactions.

\section{Conclusions \label{conclusions}}
In this work, we have developed a new approach for analyzing topological band inversions with wavefunction-based methods and demonstrated the approach in the context of the Diffusion Monte Carlo method. Even though QMC methods are well-suited for studying the effects of electron correlation on topological materials, identifying topological changes within QMC methods has been an outstanding challenge because it can be difficult for QMC to access information about excited states and produce band structures. Here, we have introduced a new method that surmounts some of these challenges by performing a momentum-space-resolved atomic population analysis on the atomic orbitals expected to contribute to the valence and conduction bands that undergo inversion. In particular, we modeled Bi$_2$Te$_3$, a well-known topological insulator, using DMC with spin-orbit coupling and showed that, by tracking the Bi and Te $p$ character of its valence and conduction bands, we can observe clear signatures of interband charge transfer and band inversion. After benchmarking our method on bulk Bi$_2$Te$_3$, we then demonstrated its utility on monolayer Bi$_2$Te$_3$, finding that there is no band inversion between the Bi-$p$ and Te-$p$ states at the $\Gamma$-point in the absence of van der Waals interactions.  Conveniently, our method is a relatively simple and cheap approach that only involves post-processing the 1-body reduced density matrices typically produced by spin-orbit DMC calculations. 

Although this first illustration of our method was made for a relatively weakly correlated material for benchmarking purposes, we expect this work will have the greatest impact on the study of strongly correlated topological insulators, such as magnetic MnBi$_2$Te$_4$, Cr-doped (Bi,Sb)$_2$Te$_3$ films, and TI heterostructures \cite{liu_2023}.  In many such materials, electron correlation can compete with topological effects, leading to subtle and non-trivial band topologies. Indeed, in some cases, correlation could widen band gaps, accommodating the emergence of nontrivial topologies. Our technique can probe whether some of these correlated materials are in fact 
topological insulators by tracking the emergence of band inversions, as a necessary, but perhaps not generally sufficient precondition. Indeed, much as QMC is often used to determine Hubbard $U$ values for the subsequent DFT modeling of quantum materials \cite{saritas_2018, wines_2022, wines_2023}, our method could similarly be employed to first determine whether a strongly correlated material is in fact an insulator that contains SOC-induced band inversions before follow-up DMC-benchmarked DFT calculations that can more precisely compute the $\mathbb{Z}_2$ invariant itself. We thus look forward to the use of our new method  - either alone or in combination with complementary or more computationally-efficient methods - for providing fresh insights into the growing class of correlated topological materials.

\section{Data Availability}

All codes, scripts, and data needed to reproduce the results in this paper are available upon request. 

\section{Acknowledgements \label{ack}}
The authors thank Daniel Staros and Paul Kent for fruitful conversations. A.L. (implementation, data analysis, writing), C.M.(code development, writing), J.A. ( mentorship, writing),  B.R. (mentorship, writing), and J.K. (concept, mentorship, analysis, writing) were supported by the U.S. Department of Energy, Office of Science, Basic Energy Sciences, Materials Sciences and Engineering Division, as part of the Computational Materials Sciences Program and Center for Predictive Simulation of Functional Materials.

An award of computer time was provided by the Innovative and Novel Computational Impact on Theory and Experiment (INCITE) program. This research used resources of the Oak Ridge Leadership Computing Facility, which is a DOE Office of Science User Facility supported under Contract No. DE-AC05-00OR22725. This research used resources of the National Energy Research Scientific Computing Center (NERSC), a U.S. Department of Energy Office of Science User Facility located at Lawrence Berkeley National Laboratory, operated under Contract No. DE-AC02-05CH11231.

Sandia National Laboratories is a multi-mission laboratory managed and operated by National Technology Engineering Solutions of Sandia, LLC (NTESS), a wholly owned subsidiary of Honeywell International Inc., for the U.S. Department of Energy’s National Nuclear Security Administration (DOE/NNSA) under contract DE-NA0003525. This written work is authored by an employee of NTESS. The employee, not NTESS, owns the right, title and interest in and to the written work and is responsible for its contents. Any subjective views or opinions that might be expressed in the written work do not necessarily represent the views of the U.S. Government. The publisher acknowledges that the U.S. Government retains a non-exclusive, paid-up, irrevocable, world-wide license to publish or reproduce the published form of this written work or allow others to do so, for U.S. Government purposes. The DOE will provide public access to results of federally sponsored research in accordance with the DOE Public Access Plan.

 All authors declare no financial or non-financial competing interests.

\nocite{apsrev42Control}
\bibliographystyle{apsrev4-2}
\bibliography{ref}{}


\end{document}